# PointGAT: A quantum chemical property prediction model integrating graph attention and 3D geometry

Rong Zhang[1*], Rongqing Yuan[2*], Boxue Tian[1#]

[1]MOE Key Laboratory of Bioinformatics, State Key Laboratory of Molecular Oncology, School of Pharmaceutical Sciences, Tsinghua University, Beijing, 100084, China

[2]Department of Chemistry, Tsinghua University, Beijing, 100084, China.

*These authors contribute equally to this work.

#To whom correspondence should be addressed:

Boxue Tian: boxuetian@mail.tsinghua.edu.cn

# Abstract

Predicting quantum chemical properties is a fundamental challenge for computational chemistry. While the development of graph neural networks has advanced molecular representation learning and property prediction, their performance could be further enhanced by incorporating 3D structural geometry into 2D molecular graph representation. In this study, we introduce the PointGAT model for quantum molecular property prediction, which integrates 3D molecular coordinates with graph-attention modeling. Comparison with other current models in molecular prediction tasks showed that PointGAT could provide higher predictive accuracy in various benchmark datasets from MoleculeNet, including ESOL, FreeSolv, Lipop, HIV, and 10 out of 12 tasks of the QM9 dataset. To further examine PointGAT prediction of quantum mechanical (QM) energies, we constructed a C10 dataset comprising 11,841 charged and chiral carbocation intermediates with QM energies calculated at the DM21/6-31G*//B3LYP/6-31G* levels. Notably, PointGAT achieved an $R^2$ value of 0.950 and an MAE of 1.616 kcal/mol, outperforming other models. Additional ablation studies indicated that incorporating molecular geometry into the model resulted in markedly higher predictive accuracy, reducing the MAE value from 1.802 kcal/mol to 1.616 kcal/mol. Moreover, visualization of PointGAT atomic attention weights suggested its predictions were interpretable. Findings in this study support the application of PointGAT as a powerful and versatile tool for quantum chemical property prediction that can facilitate high-accuracy modeling for fundamental exploration of chemical space as well as drug design and molecular engineering.

## Introduction

The quantum chemical properties of compounds are closely tied to their structures[1, 2], an accurate understanding of which is essential for molecular design and synthesis. Computational chemists employ quantum mechanical (QM) methods to quantitatively estimate chemical properties using 3D structural coordinates as input[1, 3-5]. However, QM methods such as density functional theory (DFT)[6] are resource-intensive, which has subsequently limited their application in large datasets. Therefore, the development of computationally efficient methods for predicting quantum mechanical properties is necessary for both chemistry and pharmaceutical research. Quantum chemical property prediction is a subtask of molecular property prediction, which has attracted considerable interest due to recent advances in machine learning[7-9]. Conventional machine learning[8, 10] properties models use manual molecular descriptors as input, while deep learning[11] models automatically learn features from data rather than using manual features, which is simpler, more convenient, and facilitates efficient research. Consequently, deep learning methods are increasingly applied to molecular property prediction[12-15] and drug design[16-20] tasks.

Molecular prediction models can be generally divided into four categories based on their use of 1D sequences[18, 21-23]; 2D structures[24, 25]; 3D structures[26, 27]; or molecular descriptors[28, 29]. 1D molecular representations, like SMILES (Simplified Molecular Input Line Entry System)[30] or Extended Connectivity Fingerprints (ECFP)[31], are linear encodings presenting molecular structures in strings, indicating atom and connection information. However, they lack 3D structural details. 2D molecular representations are usually 2D graphs that depict the planar structure of molecules, including the connections between atoms, the types of bonds, and their arrangements. 3D molecular representations offer insights of molecular geometry, providing atomic coordinates, bond angles, and bond lengths. Molecular structure is crucial in fields like molecular simulation, drug design, and molecular docking, as they provide detailed information about intermolecular interactions. Molecular descriptors are numeric representations for quantifying properties (e.g., molecular weight) and can be computed from 1D, 2D, or 3D representations. In particular, 2D structure-based graph neural networks (GNNs)[32, 33] have remarkably improved molecular representation learning. In GNN models[34-36], molecules are represented as 2D molecular graphs[32], with atoms and chemical bonds serving as nodes and edges[37], respectively. Among them, the graph attention (GAT) mechanism-based model, Attentive FP[34], has shown state-of-the-art performance in molecular property prediction benchmarks. Therefore, most existing models treat molecules as 2D graphs and rarely consider 3D geometric features. More recently, models such as GEM[19] and Uni-Mol[20], which are based on pre-training strategies and embedding geometric information into deep learning architectures, have also shown high performance in molecular property prediction benchmarks, suggesting that 3D molecular features can help improve the performance of molecular property prediction models. Therefore, the neural network architecture that integrates the representation of molecular graphs and geometric representations is a promising direction for molecular representation learning.

The relative QM energy of terpenoid carbocation intermediates determines their reactivity, which is key to understanding the complex mechanisms of terpenoid synthases (TPSs)[38]. Tantillo and colleagues have published extensive studies of TPS-mediated carbocation rearrangements [39-42]. Additionally, relative QM energy is a vital descriptor for characterizing reaction pathways[43, 44], which is an essential step in mapping whole reaction networks of terpene carbocation intermediates. The exploration of terpenoid chemical space[43] is essential for terpenoid scaffold-based drug design, since terpenoids are vastly abundant, constituting almost one-third of all compounds in the Natural Products Dictionary[45], and characteristically exhibit remarkable structural diversity with a wide range of biological activities[46]. Therefore, in this study, in addition to exploring the performance of molecular property prediction models in publicly available benchmark datasets, we also construct a dataset of terpene carbocation intermediates and calculate their relative energies to evaluate the accuracy of different models in quantum chemical property prediction tasks.

Here, we develop and test the PointGAT tool, which uses GAT and 3D coordinates to predict quantum chemical properties. PointGAT has two key feature-extraction models, including one for obtaining graph-level features through modified Attentive FP[34], while the other models geometric attributes using a point cloud methodology[47]. We evaluate PointGAT in several prediction tasks and datasets, including MoleculeNet[48] and the C10 dataset of monoterpene carbocations constructed in this study. We found that PointGAT can provide higher accuracy than the baseline model, Attentive FP[34], in both the MoleculeNet and C10 datasets. We also conducted an ablation study with the C10 dataset to examine the relative contribution of each PointGAT component to its accuracy. These results collectively show that PointGAT can effectively predict quantum chemical properties, facilitating fundamental discovery in chemical space as well as novel drug design while addressing major challenges of quantum mechanical property prediction.

Our main contributions are listed below:

- We introduce PointGAT, a quantum chemical property prediction model that learns molecular representation through GAT and 3D molecular geometry.
- PointGAT benchmarking was performed with five subsets of MoleculeNet, including 3 physicochemical datasets (ESOL, FreeSolv, and Lipop), and the QM9[49] dataset for regression tasks, and the HIV dataset for classification.
- We constructed a C10 dataset of monoterpene carbocation intermediates for QM energy prediction and compared the accuracy of PointGAT with other machine-learning models.

## Results and discussion

**The PointGAT framework**

PointGAT comprises three modules to incorporate molecular geometry features into a graph attention neural network (Fig. 1). The first module is graph-based, with a core design similar to Attentive FP[34] which uses dual sets of attention layers to extract features from the molecular graph. One set (comprising k layers) is dedicated to atom embedding, while the other set (consisting of t layers) focuses on full-molecule embedding. To improve accuracy, we incorporated residual connections between the atom- and molecule-level embedding. The second module is a point cloud module that extracts 3D geometric features from the molecular point clouds. We utilized a partial configuration from the PointNet[50] architecture, which consists of a section of dense feature extraction layers, followed by a max pooling operation, and finally, a densely connected layer. In this module, point clouds containing n points serve as inputs. Each point, $i$, is characterized by its relative coordinates ($x_i$, $y_i$, $z_i$) along with three additional features: charge, radius, and ring information. The third module, referred to as the output module, consists of two fully connected layers. These layers serve to combine the vectors derived from the graph-based and point cloud-based modules, ultimately generating the final predictions for molecular properties.

**PointGAT prediction accuracy with the MoleculeNet benchmarking dataset**

To assess the accuracy of PointGAT molecular property predictions, we selected five representative subsets from the MoleculeNet benchmark dataset[48], which was constructed for rigorous evaluation of molecular property prediction models. The first prediction task is ability classification for inhibiting HIV replication. We found that PointGAT provided the highest ROAUC (0.849) among all models tested with this dataset (Table 1). We then undertook three physical chemistry regression tasks, including water solubility (ESOL), solvation-free energy (FreeSolv), and lipophilicity (Lipop). PointGAT had the lowest root mean square error (RMSE) thus far reported in benchmarking for these tasks with this dataset (Table 1). Optimal physical and chemical properties are crucial for effective drugs, and precise computational prediction of these traits can significantly reduce drug development costs. Side-by-side comparison with Attentive FP showed that PointGAT had lower RMSE values for ESOL (0.487 vs 0.503), FreeSolv (0.676 vs 0.736) and Lipop (0.567 vs 0.578). In light of these results, we then compared the accuracy of PointGAT with that of other current models in a quantum mechanical (QM) property prediction task using the QM9 subset. This task included 12 quantum properties for dataset comprising approximately 134,000 stable, organic small molecules containing up to nine heavy atoms (C, O, N, and F). In this test, PointGAT again outperformed other models in predicting 10 of the 12 properties (Table 2).

Experiments on MoleculeNet benchmarks show that graph neural networks outperform traditional machine learning models that take ECFP as input features, demonstrating the powerful capabilities of graphs in molecular representation. In our work, we additionally adopt the 3D point cloud scheme to extract the geometric features of molecules. PointGAT captures graph-level information and spatial structure information, thereby improving its ability to predict various properties. In addition, previous research[51-53] suggests that stacking more layers on basic neural networks may not be feasible to improve performance. Therefore, we also introduced a residual connection between the atomic embedding and the molecular embedding to alleviate the problem of difficulty in training multi-layer neural networks and improve the performance of PointGAT. These model designs together enable PointGAT to achieve the best performance on these benchmarks.

**Construction of the C10 terpenoid carbocation Dataset**

To further investigate the accuracy of PointGAT QM energy prediction, we assembled a dataset of theoretical monoterpenoid carbocation intermediates (i.e., the C10 dataset; see Fig. 2a), including carbocations with similar two-dimensional planar structures but different three-dimensional spatial configurations (Fig. 2b). The C10 carbocation dataset was initially generated in our prior work[43] aimed at constructing a comprehensive chemical space for monoterpene carbocations. Briefly, in that previous work, we enumerated 18,758 possible monoterpene carbocations with iGen and calculated the relative energies of these carbocation intermediates with PM7[54]. For the current study, we compiled a subset of the C10 dataset containing 11,841 carbocation intermediates of monoterpenes (see Fig. 2c for the dataset construction process). Geometries were optimized at B3LYP/6-31G*, followed by single point energy calculations at the DM21/6-31G* theoretical level[55]. We performed additional benchmarking experiments using a simpler and easier-to-calculate carbocation intermediate system with 6 carbon atoms (Fig. S1), in which DM21 displayed comparable accuracy to the CCSDT (Coupled Cluster with Single, Double, and Triple Excitations) method but had a lower computational burden (Fig. 3; Table S1)[56]. It warrants mentioning that optimization with B3LYP resulted in identifying 6917 as redundant structures that undergo structural rearrangements, which decreased the total catalog of carbocations from 18758 to 11,841.

The 11,841 carbocations were clustered into 75 skeletons (Fig. S2; see Fig. S3 for energy distributions), with these skeletons primarily consisting of molecules containing one or two rings (Fig. 2d). In addition, 48.5% (5,752 out of 11,841) of the carbocation intermediates contain 5-member rings as their largest ring types (Fig. 2e). We found that the QM energies of 11,841 intermediates ranged from -44.08 to 32.56 kcal/mol, with an average of -9.84 kcal/mol. We therefore selected 3981 intermediates that had values <-15.00kcal/mol for the low energy group and the 3776 intermediates with values >-5.00kcal/mol for the high energy group, excluding all others in the intermediate range. We then examined the distribution of high- and low-energy intermediates among primary, secondary, and tertiary carbocations (Fig. 2f), which revealed 450 intermediates in the primary class (mean energy = -7.06kcal/mol), 4386 in the secondary class (mean energy = -5.71kcal/mol), and 2921 in the tertiary class (mean energy = -16.45kcal/mol), respectively.

The primary carbocations were the least in number, which also corresponds to their relatively lower stability. Among tertiary carbocations, low-energy intermediates were found in significantly greater abundance than high-energy intermediates. In contrast, high-energy intermediates were significantly more abundant than low-energy intermediates among secondary carbocations. Since secondary carbocations are relatively more prone to rearrangement reactions, we examined different skeletons of these low-energy secondary intermediates to identify potential candidates for future synthesis and engineering (Fig. S4).

The C10 dataset represents a small molecule library resource distinct from other neutral compounds in that it is comprised of charged chiral molecules containing only carbon and hydrogen, therefore provides an appropriate dataset for testing the capabilities of PointGAT prediction in structurally diverse modeling systems. Additionally, this library can serve as a novel benchmark for other researchers seeking to evaluate quantum chemical energy models.

**PointGAT accuracy in predicting QM energy in the C10 dataset**

Due to the conformational diversity of carbocation intermediates, the prediction of the C10 dataset is challenging from a machine intelligence perspective. We compare PointGAT with several GNN-based models implemented in DGL and conventional machine learning models with ECFP as input to predict QM energies on the C10 dataset. PointGAT achieved an MAE of 1.616 kcal/mol and an $R^2$ of 0.950 in the C10 test set (Table 3), with around 77% of the test samples exhibiting a prediction MAE within 2 kcal/mol (Fig. 4a). The distribution of predicted relative energies was highly correlated with that of actual energies (Fig. 4b-c), as we also observed in the QM energy landscape (Fig. 4d-e). Moreover, PointGAT outperformed most of the models in terms of RMSE, MAE, and $R^2$ values (RMSE = 2.543 kcal/mol; MAE = 1.616 kcal/mol; $R^2$ = 0.950; Table 4, Fig. 5). Comparison with the state-of-the-art molecular property prediction model, Uni-Mol[20], which was built on pre-training strategies, indicated that PointGAT accuracy was lower than that of Uni-Mol when using MMFF structures for 3D geometric feature extraction (MAE = 1.616 vs 1.584 kcal/mol; PointGAT vs Uni-Mol; Fig. S5a). By contrast, the incorporation of B3LYP geometries enabled PointGAT to provide highly accurate predictions for these structures, with an MAE of 1.228 kcal/mol (Fig. S5b). While PointGAT (MMFF94) provided higher accuracy in QM energy prediction than graph-based molecular representation models, its performance is slightly inferior to the Uni-Mol model, which is built through pre-training with 209M molecular conformations. However, the tradeoff between substantially higher computational burden of pre-trained models and a difference in MAE of 0.032 kcal/mol may be acceptable. These results collectively demonstrated that PointGAT could accurately predict the relative QM energies of carbocation intermediates in the C10 dataset, supporting its potential for prediction applications involving charged and chiral molecular configurations of complex conformations. Additionally, it provides the possibility for energy calculations of more complex sesquiterpene carbocation systems, thereby promoting the exploration of the chemical space of sesquiterpenes.

## The contribution of each module to PointGAT accuracy

To determine the relative contributions of the graph attention and 3D geometry modules to the accuracy of PointGAT predictions, we next performed ablation studies using the C10 dataset (Table 5, Table S2). Consistent with recent findings[19, 57], this analysis verified that the graph attention module was essential, and without it the MAE increased to 6.727 kcal/mol. By contrast, the removal of the point cloud-based module lowered prediction accuracy, but to a lesser extent than the graph module, indicated by an increase in MAE from 1.616 to 1.802 kcal/mol. Furthermore, to examine the influence of geometry quality, we compared PointGAT performance when using MMFF versus B3LYP geometries, which resulted in MAEs of 1.616 and 1.228 kcal/mol, respectively. These results suggested that the geometric structure features are helpful for predicting quantum chemical properties and high-precision geometry can significantly improve the accuracy of PointGAT QM energy predictions. Furthermore, removing the residual connection operation also reduced accuracy (MAE = 1.616 vs 1.653 kcal/mol), while omitting the three additional features from point clouds also decreased PointGAT performance (MAE = 1.616 vs 1.736 kcal/mol). We also observed that incorporating residual connections along with normalization not only speeds up the learning process but also enhances the model's capability to represent molecules (Table S3). These results thus show that all modules contribute to PointGAT performance, with the graph attention module playing the largest role in predictive accuracy.

## Interpretability of PointGAT

To explore the interpretability of PointGAT, we also analyzed the energy distributions and attention weights of carbocation intermediates in the C10 dataset. Subsequent 2D t-SNE projections of model features extracted prior to the output layer for all test set molecules revealed that PointGAT could clearly distinguish high-energy and low-energy intermediates in latent space after training (Fig. 6a-right) compared to before training (Fig. 6a-left). This phenomenon was particularly noteworthy considering that this was a regression task requiring the prediction of values across continuous distribution, whereas unexpectedly,

PointGAT exhibited the potential for classification functions in separating high- and low-energy molecules in the latent space. We next examined the attention weights of representative high- and low-energy intermediates from each category (Fig. 6b) and found that PointGAT assigns higher attention weights to carbocation atoms in low-energy intermediates, while assigning lower attention weights to those in high-energy intermediates, with mean normalized attention weights of 0.364 and 0.135, respectively. PointGAT assigns different weights to the positively charged carbon atoms for different energy states of the carbocation intermediates. Together, these studies demonstrate that attention weights at the atomic level have chemical implications and that different attention weights also provide insights into the model's ability to distinguish carbocation intermediates with high- and low-energy states within the feature space.

## Conclusions

In this study, we present PointGAT, an innovative framework that integrates 3D geometric features into GAT. PointGAT performance was assessed through extensive testing with five publicly available benchmark datasets, which revealed that this model could provide higher accuracy in molecular property prediction than other current algorithms. Additionally, we constructed a specialized C10 dataset comprising monoterpene carbocation intermediates, with QM energies calculated at the DM21/6-31G*//B3LYP/6-31G* levels, in which PointGAT again demonstrated higher accuracy than other available models. These cumulative results supported the analytical power and efficiency of the PointGAT toolkit for quantum chemical property prediction.

For future work, the application of PointGAT in a broader range of common molecular property prediction tasks will facilitate further validation of its accuracy in other molecular properties, further increasing its versatility. In addition, PointGAT may be applied in conjunction with different pre-trained models to combine the strengths of pre-trained models with feature extraction by PointGAT to create a unified framework capable of handling challenging molecular modeling tasks. Finally, improving the efficiency of molecular geometry extraction will expand the potential application of this model. In summary, these efforts to advance quantum chemical property prediction will pave the way for both fundamental discoveries in chemical space as well as major advances in drug development and biochemical design and engineering.

## Methods

**Dataset**

We assessed our model across six datasets (Table S3), including the ESOL, FreeSolv, Lipop, QM9, and HIV from MoleculeNet[48] and our C10 dataset. Datasets from MoleculeNet were applied for regression tasks, quantum chemistry property evaluations and classification, respectively. Among regression datasets on MoleculeNet, we specifically focused on the QM9 (Table S4) for predicting and evaluating quantum chemical properties. To evaluate the generalizability of our model across different molecular spaces, we incorporated the HIV dataset, which was specifically designed to predict the HIV inhibition activity of molecules. While other datasets are publicly available, C10 was constructed to further assess the model's predictive capability regarding quantum chemical properties. The C10 dataset comprises 11,841 carbocations. The energies of these molecules were calculated from DM21/6-31G*//B3LYP/6-31G*. Notably, each intermediate contains a total of 10 carbon atoms. This dataset had unique characteristics and properties that differed from the other datasets used in our study, making it an excellent candidate for assessing the transferability across diverse molecular spaces.

**Molecular featurization**

PointGAT included graph- and 3D geometry-level features. Graph-level features encompassed atomic and chemical bond features. Specifically, we defined eight atomic features, including atom symbol, degree, formal charge, number of hydrogens, hybridization state, chirality, ring, and chirality type; and four chemical bond features, including bond type, conjugation status, ring membership, and stereochemistry (Table 6). All these features were calculated using the RDKit[58]. The 3D geometry of a molecule was treated as a point cloud in space. To address the challenges posed by translation invariance, we utilized relative coordinates to represent each atom's position with respect to a reference point. Additionally, we employed max pooling operations to ensure that the model was invariant to rotation, capturing the most significant features regardless of the molecule's orientation. Each atom was represented by a set of six features, comprising its 3D coordinates, van der Waals radius, atomic charge, and ring membership information (Table 7). In addition, we also utilized the ECFP obtained from the RDKit package as input for conventional machine learning.

**Graph attention neural network**

The core design of our graph attention module is similar to Attentive FP. Attentive FP is a variant of GNN incorporating both atomic- and molecular-level attention mechanisms, which uses dual sets of attention layers to extract features from the molecular graph. And it consists of two key phases: the message-passing[59] phase and the readout phase[59]. In the message-passing phase, the model is responsible for aggregating information from neighboring nodes to update the representation of each node in the graph. This process ensures that each node incorporates relevant information from its neighbors, facilitating the learning of complex relationships within the molecular structure. The formulas for the message-passing phase are as follows:

$$m^{t+1} = \sum_{u \in N(v)} M_t(h_v^t, h_u^t) \tag{1}$$

where $v$ is the node and $N(v)$ is the neighborhood nodes connecting to $v$. $h_v^t$ and $h_u^t$ are the $t$ layers of the hidden state of $v$ and its neighbor $u$, respectively. In the message passing phase, through message function $M_t$, the information of the neighbors of the target node is aggregated to obtain the neighbor message vector $m^{t+1}$. The readout phase is used to compute the features of the entire graph by summarizing the information learned from individual nodes and their interactions. This global graph-level representation is crucial for making predictions about the overall properties of the molecule. The formulas for the readout phase are as follows:

$$h_v^{t+1} = GRU^t(h_v^t, m_v^{t+1}) \qquad (2)$$

The output attention context is fed into the GRU (gated recurrent unit) together with the target atom's current state vector $h_v^t$, producing the updated state vector $h_v^{t+1}$ of atom $v$.

The core idea of employing the attention mechanism in graphs is to create a context vector for the target node by concentrating on its neighboring nodes and local environment. This process involves three main steps: (1) alignment, (2) weighting, and (3) context, as explained below:

Alignment:

$$e_{vu} = leaky\_relu(W \cdot [h_v, h_u]) \qquad (3)$$

Weighting:

$$a_{vu} = softmax(e_{vu}) = \frac{\exp(e_{vu})}{\sum_{\in N(v)} \exp(e_{vu})} \qquad (4)$$

Context:

$$C_v = elu\left(\sum_{u \in N(v)} a_{vu} \cdot W \cdot h_u\right) \qquad (5)$$

$v$ is the target node, $h_v$ is the state vector of node $v$, and $h_u$ is the state vector of node $u$ (neighbor atom). In the alignment step, $[h_v, h_u]$ combines the state vectors of the target node and a neighboring node, followed by a linear transformation using a trainable weight matrix $W$. $e_{vu}$ represents the alignment result for each pair of target-neighbor node. In the weighting operation, $e_{vu}$ is normalized using the softmax function across neighboring nodes to yield $a_{vu}$, representing the weight of neighbor node $u$ to target node $v$. In the context operation, the state vectors of neighbor nodes, represented by $h_u$, undergo a linear transformation. This is followed by a weighted summation and a non-linear activation function, yielding $C_v$, which is the context vector for target node $v$.

**PointNet**

PointNet[50] is a deep learning model consisting of a 3D dataset composed of multiple points, which is commonly used to describe the geometric information of an object. The PointNet model has four layers: commencing with the input layer, the model ingests point cloud data where individual points are encoded with their 3D coordinates and additional features. Following that, in the feature extraction layer, a multi-layer perceptron (MLP) is typically utilized to process and consolidate the characteristics of each point within the cloud. Next comes the global feature layer, where PointNet incorporates a symmetric function, such as maximum pooling or average pooling, to amalgamate the features derived from all points within the point cloud. Ultimately, in the classification/regression layer, the global features obtained in the prior stage are leveraged for classification or regression tasks. One of the key advantages of the PointNet model is its ability to handle unordered point cloud data. This is achieved through the use of symmetric functions in the global feature layer, making the model permutation invariant. As a result, the PointNet model can generate consistent outputs regardless of the arrangement or rotation of the same point cloud. This rotation and permutation invariance make PointNet particularly well-suited for various point cloud-related tasks, such as point cloud classification[60], segmentation[61], object detection[62], and pose estimation[63].

**Residual connection**

As the depth of the network increased, gradient vanishing and exploding took place. To alleviate gradient-related problems, the residual connection[51, 64], providing another path for data to reach the latter parts of the neural network by skipping some layers, is widely used. Consider a series of layers, layer $i$ to layer $i + n$, and let $F$ be the function represented by these layers. Denote the input for layer $i$ by $x$. In the traditional feedforward setting, $x$ will simply go through these layers one by one, and the outcome $y$ of layer $i + n$ is $F(x)$. The residual connection that bypasses these layers typically works as follows:

$$y = F(x) + x \tag{6}$$

the residual connection expresses the output as a linear superposition of the input and a nonlinear transformation of the input.

**Training strategies**

All the data sets from MoleculeNet were experimented according to the data division standard of Attentive FP, and the C10 data set was randomly divided according to 8:1:1 (Fig. S6). Three independent runs with different random seeds to train all the models. For regression and classification tasks, we used mean-square error (MSE) and cross-entropy as loss functions, respectively. An early stop was applied to avoid overfitting. Adam optimizer[65] was used for gradient descent optimization. Three metrics (mean-absolute-error (MAE), root-mean-square error (RMSE), and coefficient of determination ($R^2$)) were used to evaluate our model for regression tasks. One classification task was evaluated by AUC (area under the ROC (receiver operating characteristic) curve). The abscissa of the ROC curve is FPR (False positive rate), and the ordinate is the true positive rate TPR (True positive rate).

$$MAE = \frac{1}{m} \sum_i^m |y_i - \hat{y_i}| \tag{7}$$

$$RMSE = \sqrt{\frac{1}{m} \sum_i^m (y_i - \hat{y}_i)^2} \tag{8}$$

$$R^2 = \frac{\sum_i^m (y_i - \hat{y}_i)^2}{\sum_i^m (y_i - \bar{y}_i)^2} \tag{9}$$

$$TPR = \frac{TP}{TP+FN} \tag{10}$$

$$FPR = \frac{FP}{TN+FP} \tag{11}$$

**Acknowledgments**

This work was supported by the Tsinghua University Initiative Scientific Research Program ((No.20221080048, No.20231080030), the Tsinghua-Peking University Center for Life Sciences (No.20111770319)

**Author Contributions**

R.Z. designed the PointGAT model and performed most of the experiments, including data preprocessing, model training, and manuscript writing; R.Y. carried out DM21 energy calculations; B.T. directed the study and helped the authors write the manuscript.

**Competing interests**

There are no conflicts to declare.

**Author Information**

The authors declare no competing financial interests. Correspondence and requests for materials should be addressed to B.T. (boxuetian@tsinghua.edu.cn).

**Code and Data availability**

We have made the code and data of this study publicly accessible at https://github.com/sevencheung2021/PointGAT, benefiting the broader research community.

Table. 1 | Comparison of performance in 4 benchmarks from MoleculeNet.

| Dataset | Metric | Splitting | Previous best ML based methods* | Previous best GNN based methods* | Attentive FP* | PointGAT |
|---|---|---|---|---|---|---|
| HIV | AUC-ROC ↑ | Scaffold | SVM: 0.792 | GC: 0.763 | 0.832 | 0.849 |
| ESOL | RMSE ↓ | Random | XGBoost: 0.99 | MPNN: 0.58 | 0.503 | 0.487 |
| FreeSolv | RMSE ↓ | Random | XGBoost: 1.74 | MPNN: 1.15 | 0.736 | 0.676 |
| Lipop | RMSE ↓ | Random | XGBoost: 0.799 | GC: 0.655 | 0.578 | 0.567 |

The values of the starred (*) models are taken from ref 30

Table. 2 | MAE values in different tasks of QM9 benchmark. (↓ indicates lower/better MAE)

| Task | Unit | GC | MPNN | DTNN* | Attentive FP* | PointGAT |
|---|---|---|---|---|---|---|
| mu | D | 0.583 | 0.358 | **0.244** | 0.451 | 0.434 |
| alpha | $b^3$ | 1.370 | 0.890 | 0.950 | 0.492 | **0.377** |
| HOMO | Hartree | 0.00716 | 0.00541 | 0.00388 | 0.00358 | **0.00335** |
| LUMO | Hartree | 0.00921 | 0.00623 | 0.00513 | 0.00415 | **0.00370** |
| gap | Hartree | 0.01120 | 0.00820 | 0.00660 | 0.00528 | **0.00486** |
| R2 | $b^2$ | 35.900 | 28.500 | **17.000** | 26.839 | 25.433 |
| ZPVE | Hartree | 0.00299 | 0.00216 | 0.00172 | 0.00120 | **0.00053** |
| U0 | Hartree | 3.410 | 2.050 | 2.430 | 0.898 | **0.441** |
| U | Hartree | 3.410 | 2.000 | 2.430 | 0.893 | **0.434** |
| H | Hartree | 3.410 | 2.020 | 2.430 | 0.893 | **0.434** |
| G | Hartree | 3.410 | 2.020 | 2.430 | 0.893 | **0.434** |
| Cv | cal/mol/K | 0.650 | 0.420 | 0.270 | 0.252 | **0.177** |

The values of the starred (*) models are taken from ref 30 and 35.

Table. 3 | PointGAT performance in the C10 dataset.

| Dataset | RMSE (kcal/mol↓) | MAE (kcal/mol↓) | $R^2$ (↑) |
|---|---|---|---|
| Test | 2.490±0.0003 | 1.616±0.0003 | 0.950±0.0010 |
| Valid | 2.653±0.0002 | 1.629±0.0002 | 0.945±0.0007 |
| Train | 1.406±0.0019 | 1.050±0.0013 | 0.985±0.0010 |

Table. 4 | Comparison of accuracy between PointGAT and other methods in the C10 test set

| Models | RMSE (kcal/mol ↓) | MAE (kcal/mol ↓) | $R^2$ (↑) |
|---|---|---|---|
| ECFP+RF | 8.588±0.0033 | 6.704±0.0037 | 0.440±0.0335 |
| ECFP+SVM | 5.989±0.0004 | 4.971±0.0005 | 0.728±0.0003 |
| ECFP+XGBoost | 4.183±0.0006 | 2.727±0.0005 | 0.867±0.0031 |
| GAT | 3.634±0.0006 | 2.641±0.0003 | 0.900±0.0025 |
| GCN | 3.504±0.0031 | 2.573±0.0032 | 0.906±0.0128 |
| MPNN | 2.803±0.0010 | 1.888±0.0007 | 0.940±0.0033 |
| Attentive FP | 2.697±0.0009 | 1.832±0.0004 | 0.945±0.0029 |
| Uni-Mol | 2.299±0.0200 | 1.584±0.0021 | 0.961±0.0045 |
| **PointGAT$_{MMFF}$** | **2.490**±0.0003 | **1.616**±0.0003 | **0.950**±0.0010 |
| **PointGAT$_{DFT}$** | **1.912**±0.0008 | **1.228**±0.0008 | **0.972**±0.0019 |

Table 5: Ablation studies of PointGAT in the C10 test set

| Models | MAE (kcal/mol ↓) |
|---|---|
| Without graph module | 6.727±0.0020 |
| Without point cloud module | 1.802±0.0005 |
| Without 3 additional point features | 1.736±0.0003 |
| Without residual connection | 1.653±0.0004 |
| PointGAT$_{MMFF}$ | 1.616±0.0003 |
| PointGAT$_{DFT}$ | 1.228±0.0008 |

Table 6. Atom and Bond features

| Atom features | Size | Description |
| --- | --- | --- |
| Atom type | 16 | [C, N, O, B, F, Si, P, S, Cl, As, Se, Br, Te, I, At, metal] (one-hot). |
| Formal charge | 1 | Electronic charge (integer). |
| Degree | 6 | number of covalent bonds [0,1,2,3,4,5] (one-hot) |
| Hybridization | 6 | [sp, $sp^2$, $sp^3$, $sp^3d$, $sp^3d^2$, other] (one-hot) |
| Hydrogens | 5 | [0,1,2,3,4] (number of bonded hydrogens, one-hot) |
| Chirality | 1 | [0/1] (whether the atom is chiral center, one-hot) |
| Chirality type | 2 | [R, S] (one-hot) |
| Aromaticity | 1 | [0/1] (Whether the atom is an aromatic atom) |
| Ring | 1 | [0/1] (whether the atom is in ring, one-hot) |
| **Bond features** | **Size** | **Description** |
| Bond type | 4 | [single, double, triple, aromatic] (one hot) |
| Stereo | 4 | [StereoNone, StereoAny, StereoZ, StereoE] (one-hot) |
| Ring | 1 | [0/1] (whether the bond is ring, one hot) |
| Conjugated | 1 | [0/1] (whether the bond is conjugated, one hot) |

Table. 7 | Point cloud features

| Point cloud features | Size | Description |
| --- | --- | --- |
| Relative coordinates | 3 | [x, y, z] (relative coordinates of atom, one-hot). |
| Charge | 1 | Electronic charge of atom (integer). |
| Radius | 1 | Van der Waals radius of atom (float) |
| Ring | 1 | [0/1] (whether the atom is in ring, one-hot) |

Table S1: Comparison of different QM methods for estimating relative energies of C6 carbocations.

| Method | MAE (kcal/mol) | Time spent (with single cpu) |
| --- | --- | --- |
| DM21/6-31G* | 0.92 | ~ 40 mins |
| B3LYP/6-31G* | 4.70 | < 1mins |
| CCSD(T)/cc-pVTZ | 0 | ~ 9 days |

Table S2: Ablation studies of PointGAT in the C10 training and validation set.

| Models | Training MAE (kcal/mol ↓) | Validation MAE (kcal/mol ↓) |
| --- | --- | --- |
| Without graph module | 4.262±0.0352 | 6.280±0.0004 |
| Without point cloud module | 1.355±0.0033 | 1.848±0.0003 |
| Without 3 additional point features | 1.022±0.0018 | 1.695±0.0003 |
| Without residual connection | 0.922±0.0011 | 1.662±0.0004 |
| PointGAT$_{MMFF}$ | 1.050±0.0013 | 1.629±0.0002 |
| PointGAT$_{DFT}$ | 0.624±0.0004 | 1.205±0.0005 |

Table S3: Experimental results with/without normalization.

| Models | Test RMSE (kcal/mol ↓) | Test MAE (kcal/mol ↓) | Test $R^2$ | Best epoch |
|---|---|---|---|---|
| Residual connections without layer normalization | 2.670 | 1.717 | 0.943 | 474 |
| Residual connections with layer normalization | 2.490 | 1.616 | 0.950 | 120 |

Table S4: An overview of benchmark datasets.

| Dataset | Number of compounds | Type | Descriptions |
|---|---|---|---|
| ESOL | 1128 | Regression | Molecular Water Solubility data for common small molecules. |
| FreeSolv | 643 | Regression | Hydration-free energy data for common small molecules. |
| Lipop | 4200 | Regression | Octanol/water distribution coefficient data for small common molecules. |
| QM9 | 133885 | Regression | Several quantum mechanical properties for small molecules. |
| HIV | 41913 | Classification | Experimentally data for the ability to inhibit virus HIV replication. |
| C10 | 11,841 | Regression | Relative energy data for monoterpene carbocation intermediates. |

Table S5: Details of QM9 dataset

| Task | Unit | Description |
| --- | --- | --- |
| mu | D | Dipole moment |
| alpha | $b^3$ | Isotropic polarizability |
| HOMO | Hartree | Energy of HOMO |
| LUMO | Hartree | Energy of LUMO |
| gap | Hartree | Gap (εLUMO−εHOMO) |
| R2 | $b^2$ | Electronic spatial extent |
| ZPVE | Hartree | Zero point vibrational energy |
| U0 | Hartree | Internal energy at 0 K |
| U | Hartree | Internal energy at 298.15 K |
| H | Hartree | Enthalpy at 298.15 K |
| G | Hartree | Free energy at 298.15 K |
| Cv | cal/mol/K | Heat capacity at 298.15 K |

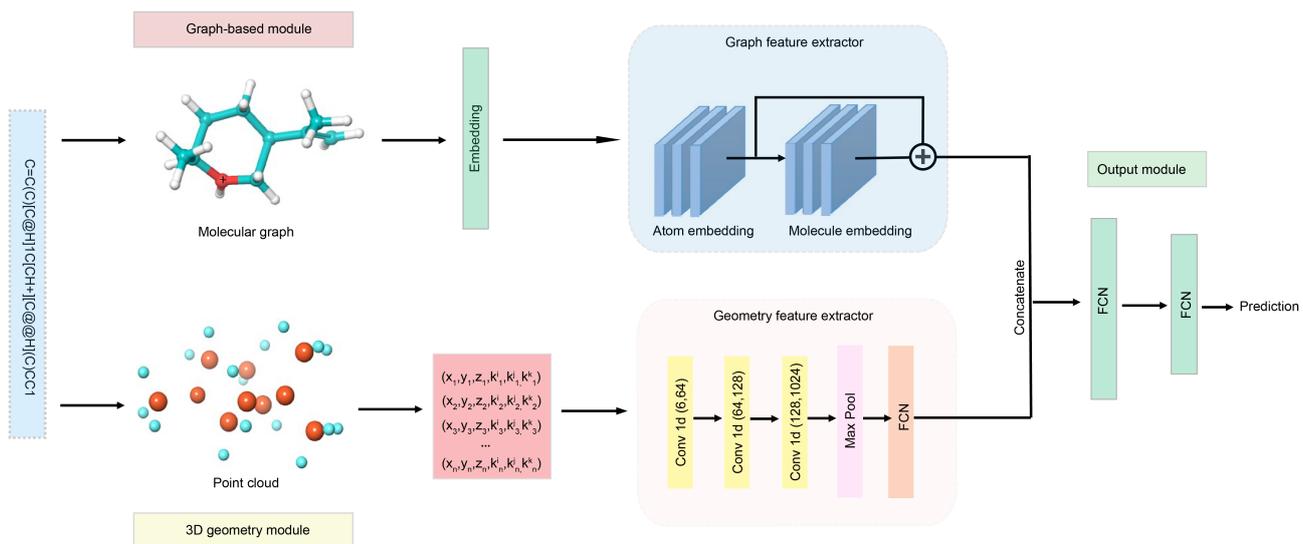

Fig. 1 | Overview of the PointGAT network architecture. Two modules are used to extract molecular features at different scales.

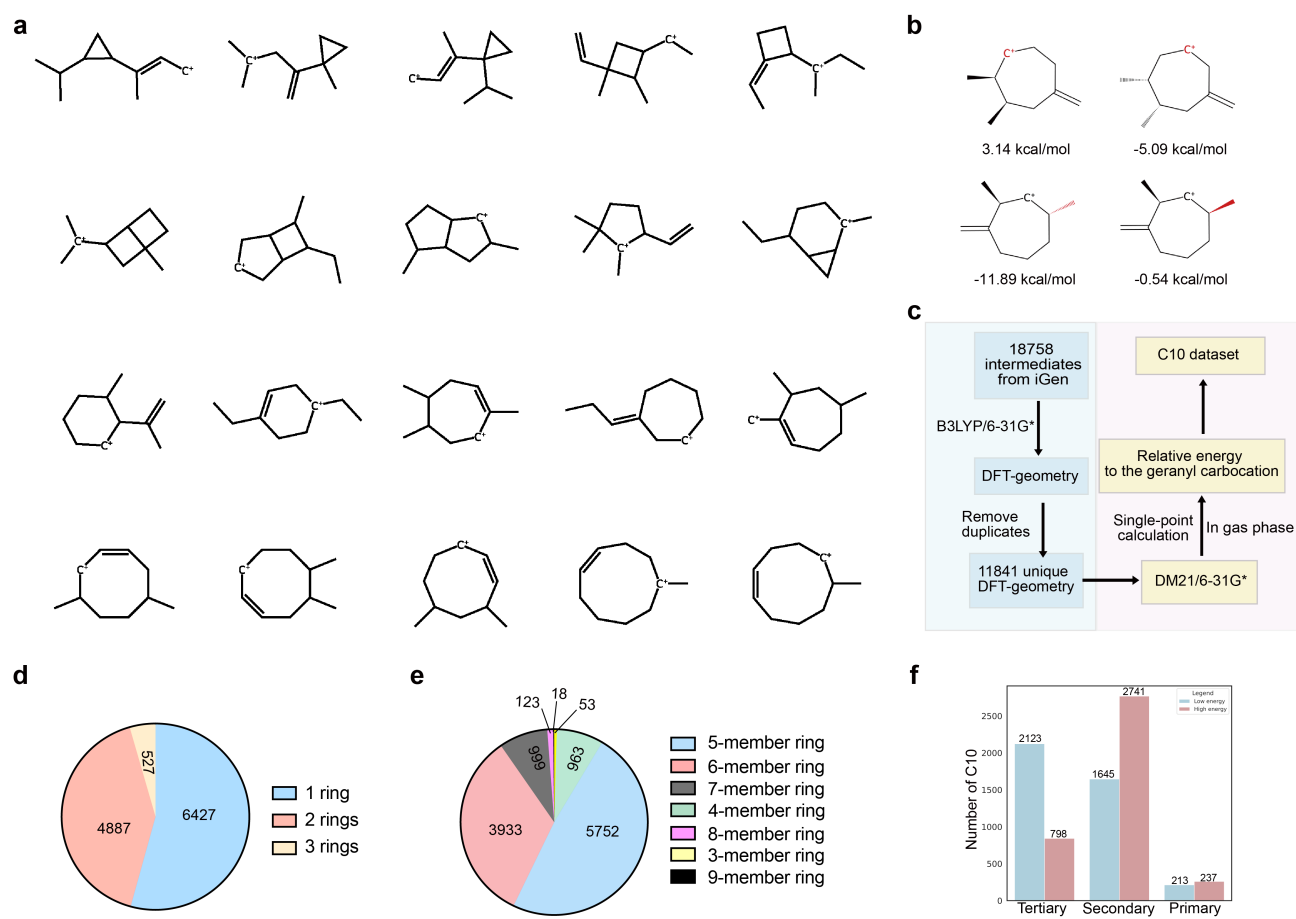

Fig. 2 | Data preparation protocol and overview of the C10 dataset. (a) Random selection of 20 carbocation intermediates from the C10 dataset. (b) Visualization of carbocations with similar 2D representations but significantly different energies. (c) Flowchart outlining the construction of the C10 dataset. (d) The count of carbocation intermediates with different numbers of rings. (e) The count of various carbocation intermediates with different ring types (the largest ring). (f) Summary of the distribution of primary, secondary, and tertiary carbocations in both high- and low-energy states.

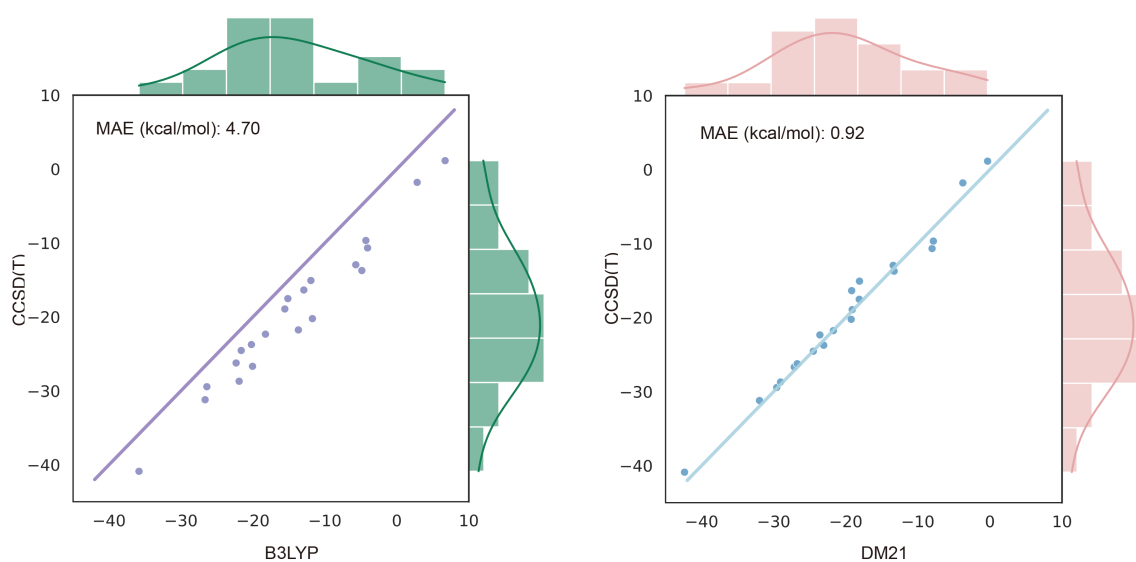

Fig. 3 | Benchmarking with the C6 dataset. A comparison of accuracy between B3LYP (left) and DM21 (right) functionals in the C6 dataset indicates that DM21 is more appropriate for QM energy calculation of carbocations.

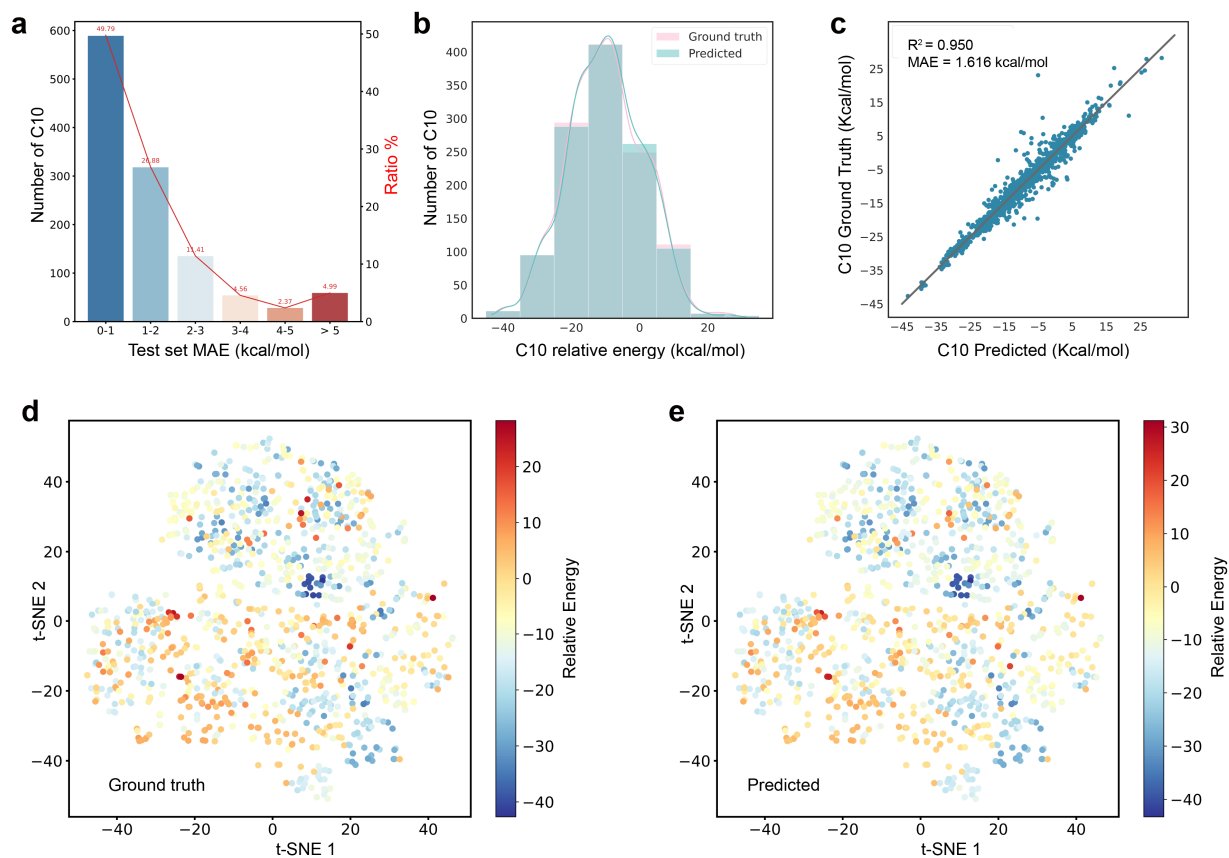

Fig. 4 | PointGAT performance with the C10 test set. (a) Histogram of MAE statistics for the C10 test set. (b) Correlation distribution and kernel density curves of predicted relative energies and ground truth for the C10 test set. (c) Scatter plot of predicted relative energies and ground truth for the C10 test set. (d) t-SNE plot of ground truth values for the C10 test set. (e) t-SNE plot of the predicted energy values for the C10 test set.

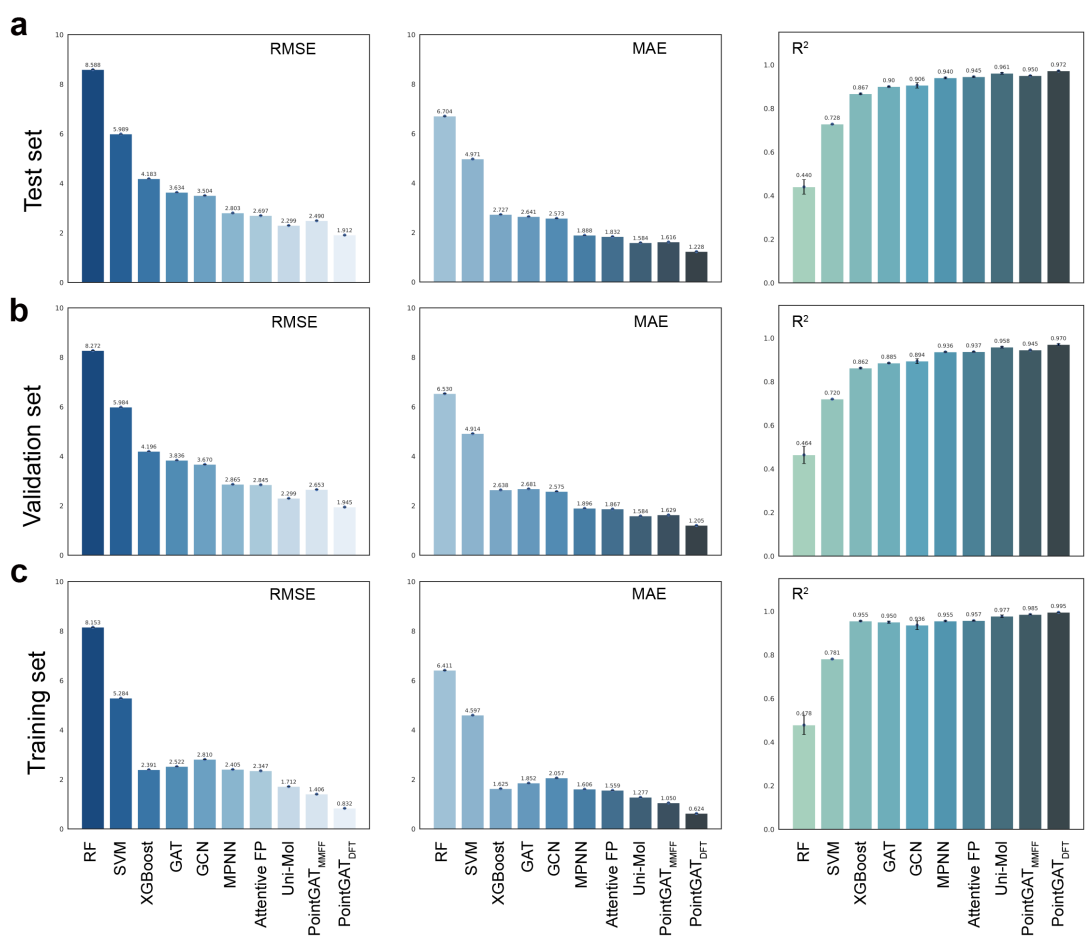

Fig. 5 | Comparison of RMSE, MAE, and $R^2$ values for the C10 dataset among different models. (a) Predictive performance in the C10 test set. (b) Predictive performance in the C10 validation set. (c) Predictive performance in the C10 training set.

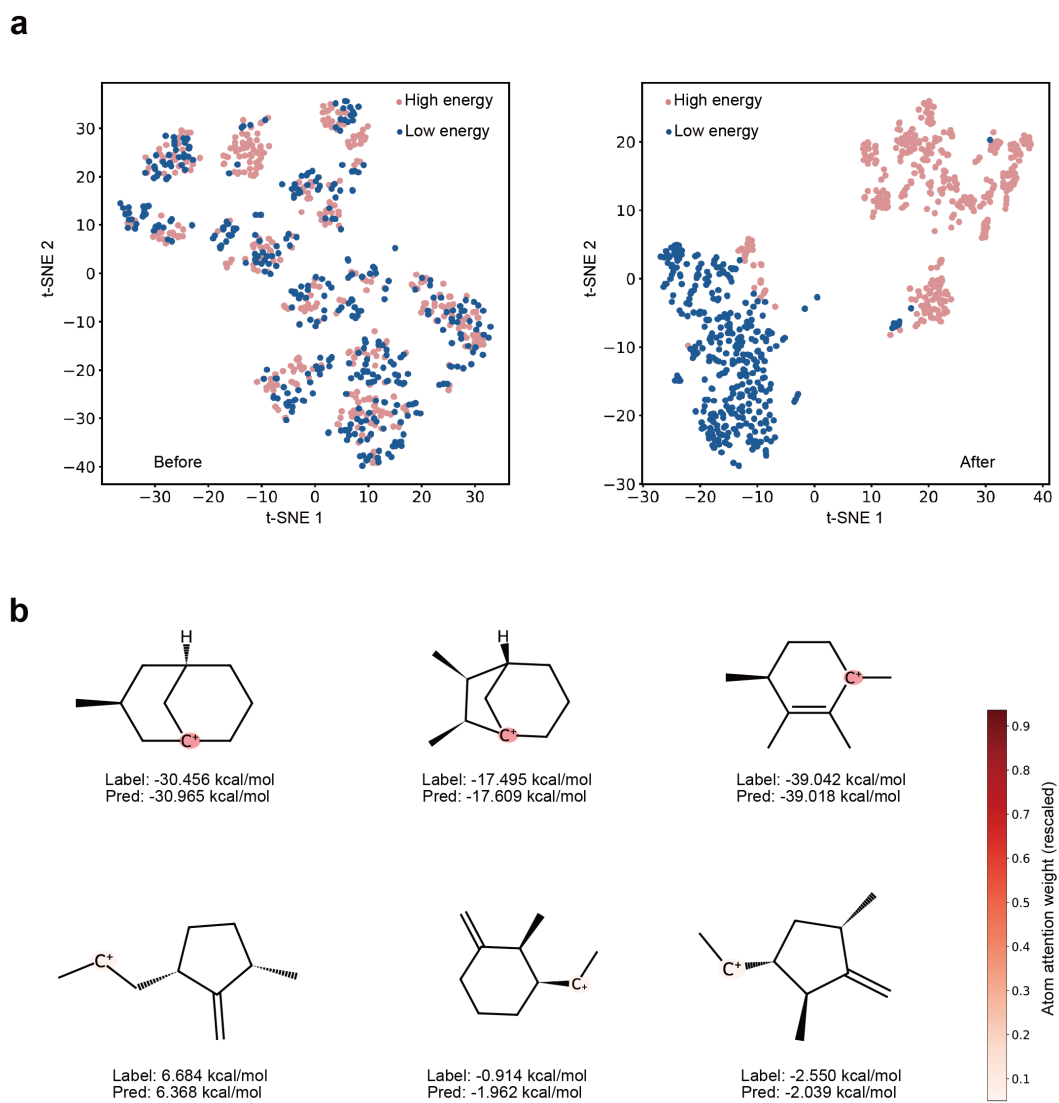

Fig. 6 | Interpretability of the PointGAT model. (a) t-SNE plot of high- and low-energy intermediate distribution in the C10 test set before PointGAT training (left), and t-SNE plot of high- and low-energy intermediate distribution in the C10 test set after PointGAT training (right), illustrating that PointGAT can distinguish high and low energy state intermediates in the latent space. (b) PointGAT assigns varying attention weights to high- and low-energy carbocation intermediates. The color bar represents attention weights.

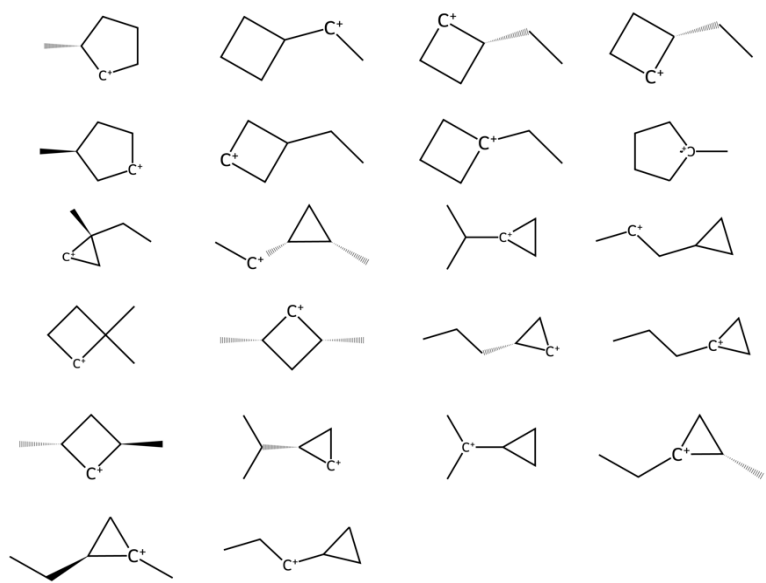

Figure S1. All carbocations of the C6 dataset.

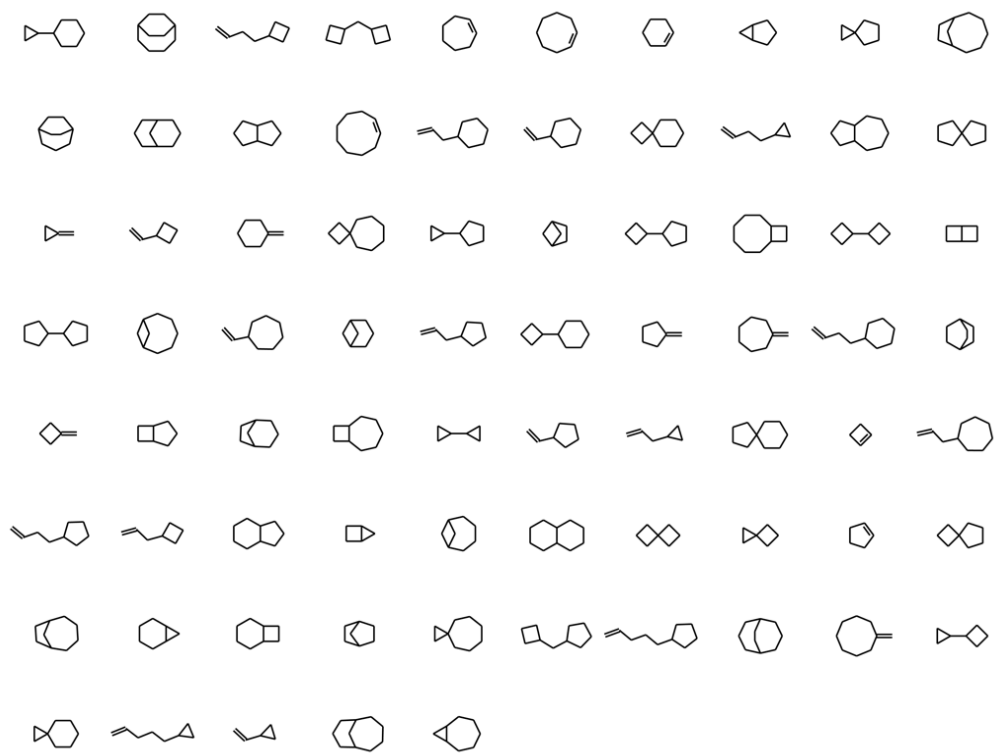

Figure S2. The skeletons of all C10 carbocations.

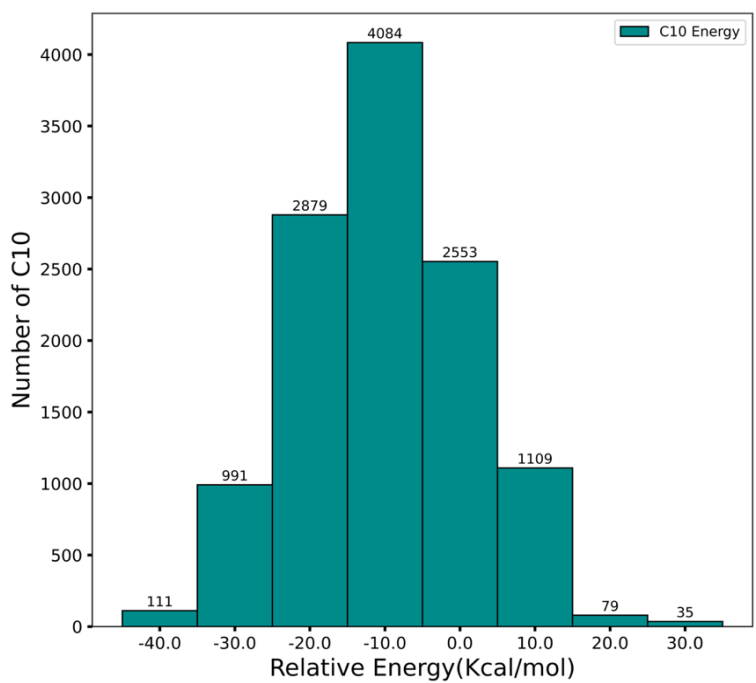

Figure S3. Energy distribution of the C10 dataset.

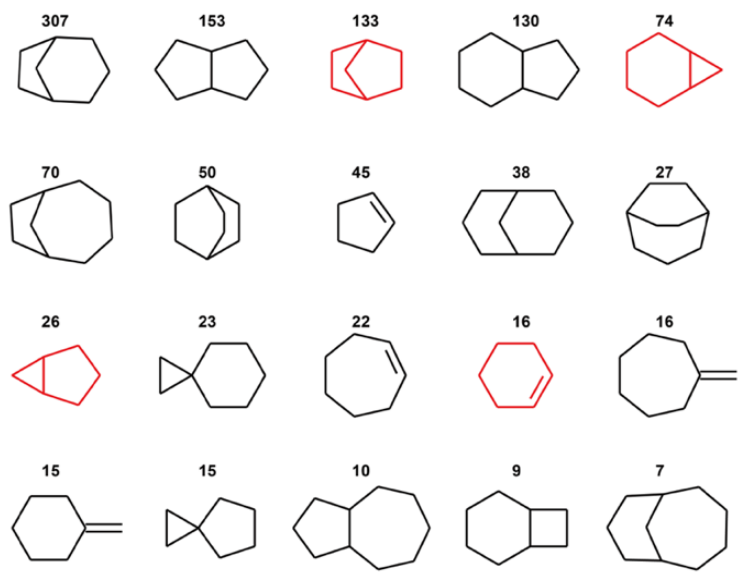

Figure S4. Top 20 skeletons of low-energy secondary carbocations. The number designates the number of intermediates from the skeleton. The skeletons highlighted in red indicate their association with products having EC numbers.

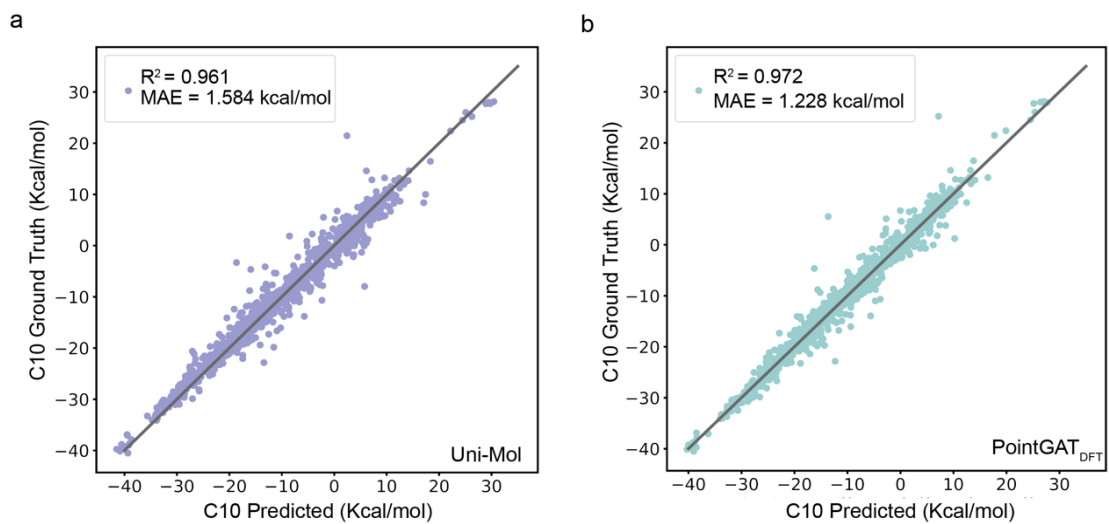

Figure S5. Scatter plot of actual relative energies vs. predicted relative energies for Uni-Mol (a) and PointGAT$_{DFT}$ (b).

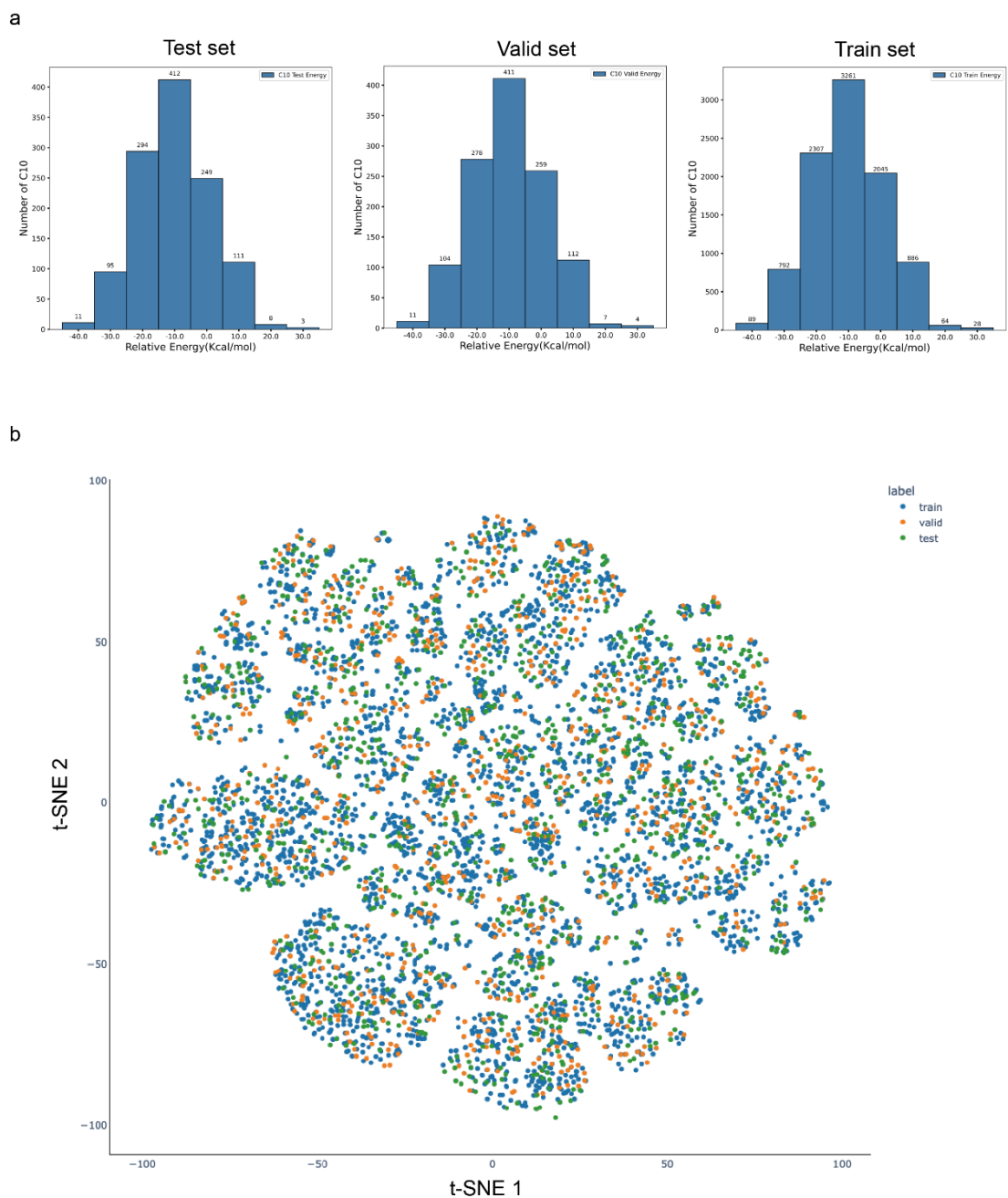

Figure S6. Data Partitioning of the C10 Dataset. (a) Energy distribution histograms for training, testing, and validation sets. (b) T-SNE visualization for the training, testing, and validation sets.